\documentstyle[11pt,epsfig]{article}
\input{epsf}
\begin{document}
\setlength{\baselineskip}{0.18in}
\newcommand{\be}{\begin{eqnarray}}
\newcommand{\ee}{\end{eqnarray}}
\newcommand{\bi}{\bibitem}
\newcommand{\lar}{\leftarrow}
\newcommand{\rar}{\rightarrow}
\newcommand{\lrar}{\leftrightarrow}
\newcommand{\mpl}{m_{Pl}^2}
\newcommand{\mplq}{m_{Pl}}
\newcommand{\rmn}{R_{\mu\nu}}
\newcommand{\gmn}{g_{\mu\nu}}

\begin{center}
\vglue .06in
{\Large \bf { 
Realistic cosmological model with dynamical cancellation of vacuum energy
  }
}
\bigskip
\\{\bf A.D. Dolgov}$^{(a)(b)(c)}$ and 
{\bf M. Kawasaki}$^{(c)}$
\\[.2in]
$^{(a)}${\it INFN, sezione di Ferrara,
Via Paradiso, 12 - 44100 Ferrara,
Italy} \\
$^{(b)}${\it ITEP, Bol. Cheremushkinskaya 25, Moscow 113259, Russia.
}  \\
$^{(c)}${\it Research Center for the Early Universe, Graduate School of 
Science, \\University of Tokyo, Tokyo 113-0033, Japan
}

\end{center}

\vspace{.3in}
\begin{abstract}

We propose a model with a compensating scalar field whose back reaction to the cosmological curvature cancels possible vacuum energy density down to the terms of the order of  the time dependent critical energy density. Thus the model simultaneously solves the mystery of the compensation of vacuum energy with the accuracy of 120 orders of magnitude  and explains existence of the observed dark energy. At an early stage the suggested cosmological model might experience exponential expansion without an additional inflaton field.

\end{abstract}

\bigskip

The problem of vacuum energy (or what is the same, cosmological constant) seems to be the most serious one in the contemporary fundamental physics.  Any reasonable estimate of the magnitude of this energy gives a non-zero result which is 50-100 orders of magnitude larger than the value allowed by astronomy and just by our existence. The potential importance of this problem was indicated in the 30th by several people after formulation of quantum field theory but more serious attitude was stimulated much later by the papers~\cite{zeld-lam}. Several mechanisms  have been since discussed to explain this mysterious and extremely precise cancellation but none has been yet satisfactory. A quite promising one is based on the idea~\cite{dolgov82} that there may exist a compensating field whose back reaction to the curvature of space-time induced by vacuum energy would cancel the latter and change the expansion regime from the de Sitter to the Friedmann one.  
A generic feature of scenarios based on this idea is that vacuum energy is never compensated down to zero but a non-compensated remnant of the order of $\mpl /t^2$ where $\mplq$ is the Planck mass and $t$ is the cosmological time. Such models predicted that there should exist in the universe a new form of matter with the energy density close to the critical one $\rho_c \sim \mpl/t^2$ with possibly quite unusual equation of state~\cite{dolgov82}. In a sense it is a good old "news" because it was found later that there indeed exists some dark energy in the universe which induces accelerated expansion~\cite{sn-high-z} and contribute about 65-70\% into total energy density~\cite{cmbr}. However, though several models with different concrete mechanisms of the compensation have been explored~\cite{dolgov85}-\cite{mr03} but none of them could satisfactory
describe realistic cosmology. In particular, the universe expansion regime was not related to the matter content of the universe. The list of the quoted above papers is probably non-complete and more references and discussions can be found in the reviews~\cite{rev-lam}.

In this work we take a scalar field $\phi$ as a compensating agent, though higher spin fields are also possible~\cite{dolgov85,dolgov91,dolgov97}. For simplicity we confine our model with the action which depends only on the curvature scalar $R$, but in higher orders in curvature the action
may depend upon the Ricci, $\rmn$, and Riemann (Riemann-Christoffel) $R_{\mu\alpha\nu\beta}$ tensors. The general action containing only first derivatives of $\phi$ 
and the curvature scalar $R$ can be written as
\be
A= \int d^4 x\sqrt{g} \left[ -\frac{\mpl}{16\pi} (R+2\Lambda) + F_1(R) + 
\frac{1}{2} F_2(R) D_\mu \phi D^\nu \phi + F_3(R) D_\mu \phi D^\mu R - U(\phi, R)
\right]
\label{A}
\ee
where the metric has the signature $(+,-,-,-)$, $D_\mu$ is the covariant derivative in this metric, 
$g=-\det[\gmn]$,
$F_j$ are some functions of the curvature scalar; in particular, the Hilbert-Einstein action term (the first one in this expression) is separated away and the function $F_1(R)$ contains possible non-linear in curvature terms; the function $U$ is a generalization of the $\phi$ potential which may also depend upon the curvature. The cosmological constant $\Lambda$ is expressed through
the vacuum energy as $\rho_{vac} = \Lambda \mpl /8\pi$. Redefining the field by 
$\phi \rar K(R) \phi$ one can always bring its kinetic term to the canonical form, 
$D^\mu \phi D^\mu \phi/2$ or annihilate the term proportional to $D_\mu R$.

In what follows we will not use the general form of the action but take a simplified version with 
$F_3 =0$, $U=U(\phi)$, $F_2 = (\mpl/R)^2$, and $F_1 = C_1 R^2$. With this choice of
$F_1$ the terms proportional to $C_1$ does not enter the trace equation for constant $R$
(see below eq. (\ref{trace})). In this form the action is similar to that considered in 
ref.~\cite{mr03} with the difference that in the quoted 
paper $F_2 = (\mpl/R)^{2n}$ with $n>3/2$. 
{\footnote {We thank Andrei Linde for informing us about this work.}}.
According to the results of ref.~\cite{mr03} the vacuum energy could indeed be compensated by the field $\phi$ but the compensation is very slow so that the expansion regime changes from the
exponential to the almost exponential one, 
$a(t) \sim \exp (\beta t^{\kappa})$, with $2/3 \leq\kappa<1$.
In this regime the energy density of the usual matter is quickly inflated away and we arrive to an empty universe not only without vacuum energy but also devoid of the usual matter.   
As we see in what follows, a smaller $n$ drastically changes the behavior of the solution

The equation of motion for the field $\phi$, as follows from the action (\ref{A}), with
$F_1 = C_1 R^2$, $F_2 = (\mpl/R)^2$, and $F_3=0$, can be written as:
\be
D_\mu\left[ D^\mu \phi\,\left(\frac{\mpl}{R}\right)^2\right] + U'(\phi) = 0 
\label{eq-phi}
\ee
In the cosmological Friedmann-Robertson-Walker (FRW) background this equation takes the form:
\be 
\ddot \phi +3H\dot \phi - 2\,\frac{\dot R}{R}\,\dot \phi + \left(\frac{R}{\mpl}\right)^2 U' (\phi) = 0 
\label{ddot-phi}
\ee
The Einstein equations are modified as
\be &&
\frac{\mpl}{8\pi}\left( \rmn - \frac{1}{2} \gmn R \right)  -
C_1 \left( 4\rmn R - \gmn R^2 \right)
 -\left( \frac{\mpl}{R}\right)^2 D_\mu\phi D_\nu \phi
  \nonumber \\
 && + \frac{1}{2}
\left( \frac{\mpl}{R}\right)^2 \left( D_\alpha \phi \right)^2 \left({\gmn } + \frac{4\rmn}{R}\right)
 -\gmn \left[ U(\phi) +\rho_{vac} \right]
\nonumber \\
&& - 2\left(\gmn D^2 -D_\mu D_\nu\right) 
 \left[ 2C_1 R - \left(\frac{\mpl}{R}\right)^2 \frac{\left( D_\alpha \phi\right)^2}{R}\right] = 
 T_{\mu\nu},
\label{ein-eq}
\ee
where $T_{\mu\nu}$ is the energy-momentum tensor of the usual matter and 
$(D_\alpha \phi)^2 \equiv D_\alpha \phi D^\alpha \phi$.

Taking trace over $(\mu-\nu)$ in this equation we obtain:
\be
-\frac{\mpl}{8\pi} R + 3 \left( \frac{\mpl}{R}\right)^2 \left( D_\alpha \phi \right)^2 - 
4\left[ U(\phi) +\rho_{vac}\right] - 
6 D^2  \left[ 2C_1 R - \left(\frac{\mpl}{R}\right)^2 \frac{\left( D_\alpha \phi\right)^2}{R}\right] = 
T^{\mu}_\mu
\label{trace}
\ee

If the usual matter is absent, i.e. $T_{\mu\nu} =0$, then the
equations (\ref{ddot-phi}) and (\ref{trace}) together with the relation
\be
R = -6 \left( 2 H^2 +\dot H \right)
\label{r-of-h}
\ee
form a complete system which allows the solution $H\sim h/t$ and $R\sim 1/t^2$. That is
instead of or after some de Sitter exponential regime of expansion we arrive to the 
Friedmann one. To see this let us look at the equation of motion of $\phi$ (\ref{ddot-phi}).
This is the equation of Newtonian dynamics with the force proportional to $R^2 U'$ and
"liquid" friction term $(3H - 2\dot R/R)$. The friction force would explode at $R=0$ if 
$\dot R \neq 0$ or, more generally, if $\dot R$ vanishes slower than $R$. This is the case if $R$ reaches zero in finite time. It implies that the field $\phi$ would stick at the
stable point of this equation at $R^2 U' = 0$ (or better to say $\phi$ would asymptotically approach the stable point). So, in equilibrium either $U' =0$ or $R = 0$. If one may neglect time derivative terms in eq. (\ref{trace}), then 
\be
R = -32\pi \left[ \rho_{vac} + U(\phi) \right]/\mpl 
\label{r-of-phi}
\ee
So at least when the motion is slow and with a proper initial value of $\phi$ the system could
evolve towards $R=0$ and hence to vanishing effective vacuum energy
$\rho_{vac}^{(eff)} = \rho_{vac} + U(\phi_0)=0$, where $\phi_0$ is the value of $\phi$ for which
the condition of vanishing of $\rho_{vac}^{(eff)}$ is realized. 

One can check that the approach to the equilibrium point asymptotically for $tm_{Pl}\gg 1$ and
$|\phi - \phi_0| \ll |U'(\phi_0)/U''(\phi_0)| \ll m_{Pl}$
can be described by the following law:
\be
\phi - \phi_0 \approx a/(m_{Pl}t^2),\,\,\,\, R\approx r/t^2,\,\,\,{\rm and}\,\,\, H\approx h/t \,
\label{asympt}
\ee
and the constant coefficients satisfy the equations:
\be
k(h)\equiv
\frac{h(2h-1)(6h-1)}{(1+3h)^2} = \frac{1}{48\pi} \left(\frac{m_{Pl}^3}{U'(\phi_0)}\right)^2
\label{h}
\ee
The other coefficients are $a = 18 (U'/m_{Pl}^3)h^2(2h-1)^2/(1+3h)$
and $r = -6h(2h-1)$. We excluded the evident solution $a=r=h=0$ which corresponds to 
$\phi = \phi_0$ when the system stays in the equilibrium point. 

The function $k(h)$ in the r.h.s. of eq.~(\ref{h}) is positive if $0<h<1/6$ or $h>1/2$. (We do not consider here the contraction regime when $h<0$.) So possible interesting solutions lie in one of these two regions, depending upon initial conditions. The maximum value of $k(h)$ in the 
interval $0<h<1/6$ is approximately 0.0243. So the solution in this interval exists if 
$|U'(\phi_0) | > 0.52 m_{Pl}^3$. 

There could be also another regime of approach to equilibrium which, at least asymptotically, agrees with equations of motion  (\ref{ddot-phi},\ref{trace},\ref{r-of-h}). In this regime
the last term in the r.h.s. of eq.~(\ref{trace}) proportional to $D^2$ becomes non-negligible
and the solution takes the form $\dot \phi \approx -U'(\phi_0) R^2 t/m_{pl}^4$.
In this case both $R$ and $\phi -\phi_0$ tend to zero faster than
$1/t^2$: $R\sim 1/t^{2+\sigma}$ and $\phi - \phi_0 \sim 1/t^{2+2\sigma}$
and the Hubble parameter decreases as $1/t^{1+\sigma}$ with $\sigma$ satisfying
$\sigma(\sigma+1) = m_{Pl}^6/(48\pi (U')^2)$. If such regime were realized the universe would reach stationary state with a constant scale factor. We assume that such a pathological situation does not exist.

One can check that there are several branches of the solution. For example the determination
of $R$ from eq.~(\ref{trace}) through $\dot \phi$ and $(U+\rho_{vac})$, in the case that the last term in the r.h.s. of (\ref{trace}) may be neglected, can be done by solution of cubic algebraic equation which has three possible roots. The choice of a certain root is dictated by initial conditions. We have found numerically that there exists a solution for which $Ht \rar const \approx 0.5$ and $R$ tends to zero somewhat faster than $1/t^2$. So the system approaches the expansion regime typical for relativistic matter. A detailed numerical study of different solutions will be presented elsewhere.

The cosmological expansion regime is determined by the value of the constant $h$ according to $a(t)\sim t^h$. The evolution of the energy density of the usual matter is governed by the covariant conservation of its energy-momentum tensor:
\be
\dot \rho_m = -3 H (\rho_m + p_m)
\label{dot-rho}
\ee
where $\rho_m$ and $p_m$ are respectively energy and pressure densities of matter. So, as is well known, the energy density of relativistic matter decreases as $\rho_{rel}\sim 1/a^4$, while that of non-relativistic evolves as $\rho_{nr} \sim 1/a^3$. Thus, if initially the energy density of matter was sub-dominant with respect to the vacuum energy and $h<1/6$ the solution presented above should change its character because at a certain moment the usual matter would start to dominate.
Probably the energy density of matter and not compensated remnant of vacuum energy would manifest tracking behavior as suggested in ref.~\cite{quint}. 
One more comment worth making here. For relativistic matter $T^\mu_\mu = 0$ and the solution found above should be valid in the presence of such matter. However even a small admixture of matter violating the condition of vanishing of $T^\mu_\mu$, either by existence of massive particles, even with $m<T$, or by quantum trace anomaly could strongly change the behavior of the vacuum solution if $\rho_m$ drops slower than $1/t^2$, i.e. for $h<2/3$

If, at some stage, the universe was dominated by relativistic matter and vacuum energy was vanishingly small then, according to equations (\ref{ddot-phi}) and (\ref{A}), $\phi$ would be zero or a constant and the cosmology would be the usual Friedmann one.  It is well known, however,
that in the course of expansion and cooling down several phase transitions could take place in
cosmological plasma. Typically in the course of phase transition vacuum energy changes and
if it was zero initially it would become non-vanishing and, if $\phi$ remained constant, the newly created $\rho_{vac}$ would ultimately dominate the total energy density, because $\rho_{vac}$ does not decrease in the course of expansion.  However, when $\rho_{vac}$ becomes cosmologically essential, the compensation mechanism, described above, should be switched-on and  the solution (\ref{asympt}) would be approached. However, in this case the initial conditions would probably demand $h>1/2$. More detailed investigation is in progress.

If the discussed here mechanism indeed operates in the real universe, then assuming that at the 
present (and somewhat earlier) time the energy density of matter is (was) equal to its observed value, $\rho_m = \mpl/18\pi t^2$ we find from eqs.~(\ref{ddot-phi},\ref{trace},\ref{r-of-h}) that it could be realized either for $U'/m_{Pl}^3 \approx 0.0045$ or $U'/m_{Pl}^3 \approx 13.5$. The simple solution (\ref{asympt}) does not describe the observed universe acceleration but it may be because it is not precise and more accurate treatment would reveal the accelerated behavior.

The model presented here may be compatible with inflationary scenario without an additional inflaton field. If initially $\rho_{vac} \neq 0$ and the feed-back (or better to say "kill-back") effect of $\phi$ is sufficiently slow (this can be achieved by initial conditions and the shape of $U(\phi)$),
there would be a period of super-luminous (inflationary) expansion. The matter might be created by the gravity itself. As was shown many years ago~\cite{grav-prod} particle production by the Planck scale gravitational field could be quite significant. 

We do not intend to present a detailed discussion of cosmology, including big bang nucleosynthesis, large scale structure formation, etc in this short paper. It should be a subject of a much longer work. We only want to demonstrate here that in the considered scenario vacuum energy is automatically canceled out with the accuracy of the order of $\rho_c\sim \mpl/t^2$ and the usual matter could survive in the course of the cosmological evolution (it was a strong drawback of previously considered models). Moreover the non-compensated remnant of the vacuum energy could be an excellent candidate for the observed today dark energy.

\bigskip

{\bf Acknowledgment} A.D. Dolgov is grateful to the Research Center for
the Early Universe of the University of Tokyo for the hospitality during
the time when this work was done.

\end{document}